\documentstyle[prd,aps,floats]{revtex}

\newcommand{\pd}{\partial}
\newcommand{\kap}{\kappa_5^2}
\newcommand{\kapp}{\kappa_5^4}
\newcommand{\lag}{{\mathcal{L}}}
\newcommand{\half}{\frac{1}{2}}
\newcommand{\jump}[1]{ \left[ #1 \right]^+_- }
\newcommand{\varlag}[1]{{\frac{\delta \lag}{\delta #1}}}
\newcommand{\delbar}{\overline{\nabla}}

\newcommand{\Rbar}{\overline{R}}
\newcommand{\Gbar}{\overline{G}}
\newcommand{\Lambar}{\overline{\Lambda}}
\newcommand{\diag}{\mbox{diag}}
\newcommand{\eps}{\varepsilon}
\newcommand{\appleq}{\raisebox{-3pt}{\makebox[14pt]{$\stackrel{\textstyle<}{\sim}$}}}

\begin{document} 


\preprint{}
\title{Cosmological expansion on a dilatonic brane-world}
\author{Andrew Mennim and Richard A. Battye}
\address{Department of Applied Mathematics and Theoretical
Physics, Centre for Mathematical Sciences, University of Cambridge, \\
Wilberforce Road, Cambridge CB3 0WA, UK} 
\maketitle

\begin{abstract}

In this paper we study brane-world scenarios with a bulk scalar field,
using a covariant formalism to obtain a 4D Einstein equation via
projection onto the brane.  We discuss, in detail, the effects of the
bulk on the brane and how the scalar field contribute to the
gravitational effects.  We also discuss choice of conformal frame and
show that the frame selected by the induced metric provides a natural
choice.  We demonstrate our formalism by applying it to cosmological
scenarios of Randall-Sundrum and Ho\v{r}ava-Witten type models.
Finally we consider the cosmology of models where the scalar field
couples non-minimally to the matter on the brane.  This gives rise to
a novel scenario where the universe expands from a finite scale factor
with an initial period of accelerated expansion, thus avoiding the
singularity and flatness problem of the standard big bang model.

\end{abstract}

\newcommand{\beq}{\begin{equation}}
\newcommand{\eeq}{\end{equation}}

\section{Introduction}

The standard Big-Bang model based on the FRW spacetime now has many
observational successes, notably the observed expansion of distant
galaxies, the synthesis of light elements during Big Bang
Nucleosynthesis (BBN) and the isotropy of the Cosmic Microwave
Background (CMB). These successes mainly probe what, from the point of
view of this paper, we shall describe as the late universe, that is,
after neutrino decoupling --- which is just before BBN. Our knowledge
of the universe before BBN is very limited and hence it is possible
for substantial modifications to our understanding of the expansion of
the universe during such an epoch.

Motivated by attempts to solve the hierarchy problem of particle
physics and to reconcile cosmology with string theory, much interest
has been generated by the idea~\cite{RubS,dim} that this Big-Bang model is
embedded inside some higher dimensional spacetime; the 4D universe
that we see being thought as a 3-brane or brane-world. In such models
the matter fields of the standard model are localized by some mechanism
to only have support on the 3-brane,  with gravity and other
undetected particles, such as scalar fields, propagating in the
directions perpendicular to the brane  or `bulk' as they have become
known. In the most of popular of these models, the
Randall-Sundrum (RS) scenario~\cite{RS1,RS2}, the brane is  a
hypersurface (or co-dimension 1 brane)  in a 5D spacetime with $Z_2$
symmetry along the extra dimension. 

In fact, RS made two simple proposals (see also ref.~\cite{early} for 
related earlier work):
the first~\cite{RS1} (RS1) has two branes of opposite tension in an
anti-de-Sitter (adS) background spacetime with $Z_2$ symmetry. In such
a model the global spacetime structure is warped, that is, the spatial
3-metric contains an exponential or warp factor which is a function of the
coordinate along the extra-dimension. They suggested that this
could possibly solve the hierarchy problem of particle physics by
relating the Planck scale on our brane (the one with negative tension)
to that on the other via an exponential
function of the distance between them.  Hence, a large hierarchy
($\sim 10^{16}$) can be generated from a relatively small dimensionless number.

In the second such model~\cite{RS2} (RS2), there is only a single
positive tension brane: it being possible to consider this as a
special case of RS1 with physical interpretation of 
the branes reversed and the negative tension
brane taken off to infinity.
In this scenario, which has an effectively infinite 
extra-dimension cut-off by the adS horizon, they showed that there exists
a normalizable gravitational zero mode coupled to the brane which
implies that to some degree the gravitational effects on the
brane are close to those predicted by Newtonian theory and Einstein's
4D General Relativity (GR), even though the spacetime has an extra
dimension. More precisely they showed that the gravitational theory
experienced by observers on the brane is a tensor theory, being
mediated by a massless spin 2 particle, with a tower of massive
Kaluza-Klein type states due to the extra dimension. Although there is
no natural solution to the hierarchy problem in this model, much
interest has been generated by the possibility of creating such a model
within a modern fundamental theory based on string or M-theory.

Interestingly, a very similar proposal already existed in the
literature: the Ho\v{r}ava and Witten (HW) theory~\cite{hw,witten}.
In this model, which is derived from the strong coupling behaviour of
$E_8\times E_8$ string theories in 10D, the $Z_2$ symmetry is natural;
it being a consequence of the orbifold structure of the related 11D
M-Theory.  Orbifolds, manifolds which are smooth except for a finite
number of points, occur naturally in realistic compactifications
of string theories since compactification on smooth manifolds cannot
generate certain aspects of the Standard Model, for example, the chiral
properties of neutrinos.  Furthermore, the fixed points of the
orbifold give natural places for the branes to live.  In the HW case
the orbifold is $S^1/Z_2$ so there are two fixed points.
This is similar to the RS1 model except that vacuum energy
contributions are generated by the effects of a scalar field with bulk
support.  Simple cosmological solutions were studied, in
refs.~\cite{Ovrut,Harvey}, although very little was learned
about the detailed dynamics of the expansion.

The existence of the gravitational zero mode coupled to the brane
suggests that the gravitational effects are close to those of 4D GR,
but there are subtle differences, particularly on small
and large scales. For example, using a linearized analysis,
it was shown in ref.~\cite{GT} that
a localized mass distribution on the positive tension brane
of RS1 will effect the gravity on the negative tension brane
in such a way as to make it appear to be like 4D Brans-Dicke (BD)
gravity.  The fact that any brane-world model must give rise to a
universe which expands in an essentially similar way to the standard
FRW universe, at least from the epoch of BBN,
implies that one can probe the gravitational dynamics of
such model at large distances using cosmology. Probing the viability of
brane-world models in which there is an extra scalar field with bulk
support using cosmology is the main motivation of this
paper, along with the possibility that such scenarios might alleviate
the many naturalness problems, such as the cosmological
constant problem,  prevalent in 4D FRW models.

A large number of papers have attempted to understand the cosmological
expansion rate in RS type\footnote{We shall refer to a RS type model
as one with just constant vacuum energy terms, as opposed to those  which
are generated by the existence of a scalar field.} brane-world models, this
being the first thing that one must get right before making any more
exotic predictions. Here, we give a brief, and no doubt incomplete,
history of the main recent developments. One of the remarkable aspects
of this area is that it has been shown that there exist spatially
homogeneous and isotropic brane solutions whose expansion can be
understood almost totally without knowledge of  the detailed 5D
solution, with only small corrections to the expansion rate
due to the effects of the bulk. Initially, it was
suggested~\cite{BDL1,CF} that a simple model for a 3-brane in a
flat 5D background would lead to a Hubble expansion rate $H\propto
\rho$, where $\rho$ is the density of the matter on the brane, in
complete conflict with the observed expansion rate ($H^2\propto \rho$
in a 4D FRW universe) and hence with BBN\@. However, this ignored the
lesson of the RS1 and RS2 models. In the simplest versions of these
two models vacuum energy, or cosmological constant (CC), contributions
were placed on the two branes along with a negative CC in the bulk to
give asymptotic adS geometry.
 Decomposing the matter on the brane into a CC plus
other matter and including a negative CC in the bulk rectified the
problem assuming the standard RS relation between the
CCs~\cite{cline,csaki1,BDL2,Flan2,shiro}.  This modification leads to
the rather perplexing result that on the negative tension brane
$H^2\propto -\rho$. It was suggested in~\cite{csaki2,lesgourges}, and
more recently in~\cite{csaki3,cline2} that an
averaging procedure {\it a la} Kaluza-Klein could rectify this
situation if one introduced a bulk scalar field to stabilize the extra
dimension via, for example, the Goldberger-Wise mechanism~\cite{GW}. Another
interesting consequence of these models is a  radiation type
contribution to the expansion rate due to the effects of the
bulk~\cite{BDL2,Flan2,shiro,V,Muk}.

Cosmological solutions  in brane-world models with a bulk scalar field
(see refs.~\cite{CR,Ull,KOP} for previous work on this subject) are
necessarily more complicated due to the extra field.  However, they are
probably more interesting since it would seem unlikely that only
gravity would propagate in the bulk. In this paper we attempt to make
some progress in understanding how the inclusion of such a field will
effect the cosmological evolution, and in particular whether the
standard FRW expansion is possible at late times. In contrast to the
case of an RS type brane-world model, we find that there is an
obstacle to a closed form solution in these models since the evolution
on the brane can be affected in an almost arbitrary way by the
dynamics of the scalar field in the bulk. In particular, it is not
possible to derive an effective equation for the evolution of the
scalar field on the brane, in terms of quantities which are only
defined on the brane. In order to make any statements about the
late-time behaviour of the expansion, therefore we are forced to make the
reasonable assumption that the value of the scalar field on the
brane is either stabilized or slow varying. We will see that the
expansion rate is related to value of the scalar field, and hence if
it is varying, one would have a model with a variable effective
gravitational constant, which is constrained by
observations to be at least slowly varying after BBN.
An interesting consequence of this approximation is that the
cosmological expansion of stabilized or slowly varying dilatonic brane
worlds is essentially that of an appropriately tuned RS type model.

Our approach to this problem will be to use the Gauss-Codacci
formalism which gives a coordinate independent description
of the dynamics on the brane in terms of effective 4D Einstein
equations~\cite{shiro}. In section \ref{sec:form} we extend the
analysis of ref.~\cite{shiro} to the situation where there is a scalar
field in the bulk.  Note that these methods are very
different to those used in most of the other 
literature~\cite{BDL1,CF,cline,csaki1,BDL2,Flan2,V,Muk} probing
cosmology on RS type brane-worlds where specific form of the metric
is used. We show how the results of these two different approaches can
be reconciled easily and, as a consequence, gain an improved
understanding of the effects of the bulk.  As we are dealing with a
theory with a scalar field, the possibility of choosing a different
conformal frame arises, as in 4D scalar-tensor gravity theories.
We discuss the possibility of conformal scalings of the metric in
section~\ref{sec:conf}. 
We show, firstly, that our frame is the preferable one on the grounds
that it is the only frame in which the energy-momentum contribution
due to the scalar field does not contain second derivatives of the
scalar field and, secondly, that the conformal scaling, although allowing
the sign of the Hubble parameter to change, does not affect the
relationship between the signs of the brane tension and the effective
Newton constant.

In section~\ref{sec:apps} we apply these methods, first, to the
simplest RS type model obtaining the results of ref.~\cite{shiro},
before applying them to the more complicated models which include a
stabilizing scalar field  potential for the size of the extra
dimension. We show that once the scalar field becomes stabilized the
4D Einstein equations become exactly those of the RS model and hence a
negative tension brane leads to $H^2\propto -\rho$ as in the case of
constant potentials.  We discuss in detail the conflict between the
results obtained by this approach and those of other analyses, which
have concluded that the standard FRW result should apply.  This raises
the important question of what to interpret as the effective 4D theory.

We then study a simple HW type model with exponential potentials in
the bulk and on the brane. Assuming that the matter on the brane is
not coupled to the scalar field, and the scalar field on the brane is
varying slowly, we show that the RS type cosmological evolution
is still valid, albeit now with an effective gravitational constant which
is varying with time. It is interesting to note that in such models
energy conservation on the brane is modified by the variation of the
scalar field giving rise to a novel scenario for the variation of the
coupling constants. If one allows the matter on the brane to be
coupled to the scalar field as well, then it is possible to achieve
accelerated expansion on the brane leading to an interesting bulk
driven inflation model even when the scalar field is stabilized.

\section{Formalism}
\label{sec:form}

\subsection{Geometrical description of the spacetime}
\label{sec:geom}

The basic method is to foliate the 5D space-time in the direction of
the extra-dimension as done in ref.~\cite{shiro}. This makes a 4+1
split of the Einstein equations in a similar way to the
standard 3+1 split used in treatments of the Cauchy initial value
problem in GR (see, for example, ref.~\cite{wald}). This geometrical
split allows us to understand the precise gravitational dynamics on
the brane without any possible problems with unphysical gauge degrees
of freedom.

More precisely, we wish to study space-times\footnote{Note that,
throughout, we will use the same sign conventions as \cite{wald,mtw}: in
particular, we use a $(-++++)$ signature for the 5D space-time
metric.} including a smooth 4D time-like hypersurface (or co-dimension
1 brane) which we can express, in
terms of the coordinates $x^a$ with $(a=0,1,2,3,4)$,  by $f(x^a)=0$, where
$f$ is a real-valued function.  Hence, we can define a space-like
normal vector $\partial_af$ and a unit vector $n_a$ parallel to it by
\beq 
n_a = \pd_a f / (\partial_bf\partial^bf)^{1/2} \,.
\eeq
One would like to extend $n_a$ to follow a geodesic congruence, that is, so
that it satisfies the geodesic equation, $n^b\nabla_bn^a=0$.
Near the brane, this will define a
smooth foliation, but it may not be possible to foliate the entire bulk
in this way.  Moreover, in cases where there is more than one brane,
a geodesic congruence which is normal to one brane may not be normal
to another.  For example, a coordinate system which is Gaussian-normal
with respect to one of the branes, that is, $n_a=(0,0,0,0,1)$
on that brane, is not necessarily Gaussian-normal with respect to 
another~\cite{CGR}.  (See also \cite{MSM} for a discussion of the
regions covered by Gaussian normal coordinates in Schwarzschild-adS space.)

One can define the metric, or (first) fundamental tensor, of the
foliation, $h_{ab}$, in terms of the normal vector $n_a$ to be its orthogonal
complement,
\beq
{h^a}_b = {\delta^a}_b - n^a n_b \,.
\eeq
As well as being the higher-dimensional manifestation of the induced
metric on the brane, this is a projection tensor which, along with the
it orthogonal complement, allows one to project other tensors
tangential on to the brane. As an example consider any vector, $V^a$, which
can be written as 
\beq
V^a={h^a}_bV^b+(n_bV^b)n_a\,.
\label{proj}
\eeq
Clearly, the first part is tangential to the brane and the other is
perpendicular. We shall see in subsequent sections that this
projection property is an important concept when to attempting to understand
brane-world since it allows one to split the dynamics up into those
tangential to the brane and those perpendicular.

The extrinsic curvature, or second fundamental tensor, is defined to be
\beq
K_{ab}=\nabla_a n_b={h_a}^c{h_b}^d\nabla_c n_d\,,
\eeq
which is symmetric by Frobenius' theorem and tangent to the
hypersurface. It can also be written in terms of Lie derivatives of
the metric with respect to the normal 
\beq
2K_{ab} = \pounds_n g_{ab} = \pounds_n h_{ab}\,,
\eeq
and its Lie derivative is given by
\beq
\pounds_n K_{ab}=K_{ac}{K_b}^c-R_{acbd}n^cn^d\,,
\eeq
where $R_{abcd}$ is the Riemann curvature tensor of the background space-time.
In a Gaussian-normal coordinate system, the extrinsic curvature
is half of the derivative of the hypersurface metric, $h_{ab}$, with respect
to the coordinate normal to the hypersurface.

Within this setup, we can use~\cite{shiro}
the Gauss-Codacci formalism to compute
effective 4D Einstein equations for the gravitational dynamics on the brane.
We will denote all curvature tensors pertaining to the hypersurface metric,
$h_{ab}$, with a bar over the letter, whereas those pertaining to
$g_{ab}$ will have no bar.
Similarly, $\delbar$ is the covariant derivative that preserves
$h_{ab}$  whereas $\nabla$ is the covariant derivative that preserves
$g_{ab}$.  
The Gauss and Codacci equations are then
\begin{eqnarray}
\label{gauss}
& \Rbar_{abcd}= {h^j}_a {h^k}_b {h^l}_c {h^m}_d R_{jklm} + 2 K_{a[c}
K_{d]b} \,, & \\
\label{codacci}
& \delbar_b {K^b}_a - \delbar_a K =  n^c {h^b}_a R_{bc} =  n^c
{h^b}_a G_{bc}\,.&
\end{eqnarray}
Following the approach of ref.~\cite{shiro}, we use the decomposition of
the 5D Riemann tensor~\cite{wald}
\beq
\label{decomp}
R_{abcd}=\frac{2}{3} \bigg( g_{a[c} R_{d]b} - g_{b[c} R_{d]a} \bigg)
         -\frac{1}{6} R g_{a[c} g_{d]b} + C_{abcd} \,,
\eeq
which we substitute into (\ref{gauss}).
Contracting appropriately gives the following expression for the 4D
effective Einstein tensor,
\beq
\label{gbar}
\Gbar_{ab}=\frac{2}{3}\left\{G_{cd}{h^c}_a{h^d}_b+\left(G_{cd}n^cn^d-
\frac{1}{4}G\right)h_{ab}\right\}+KK_{ab}-{K_a}^cK_{bc}-\frac{1}{2}
\left(K^2-K^{cd}K_{cd}\right)h_{ab}-E_{ab} \,,
\eeq
where $E_{ab}$ is the electric part of the Weyl tensor with respect to
$n_a$, given by
\beq
E_{ab}=C_{acbd} n^c n^d \,.
\eeq

We should note that at this stage everything we have derived is a
geometric identity and hence true on all of the 4D hyperspaces in the
foliation (assumed to be well-defined). In subsequent sections we will use
junction conditions to compute $K_{ab}$ close to the brane and use
this expression for the Einstein tensor to make a comparison with the
predictions for cosmological expansion of the brane and those of 4D GR
and  other gravity theories

\subsection{The action and variation}

In order to impose some dynamics on the geometrical identities of the
previous section we will derive equations of motion from a variational
principle which includes the standard Einstein-Hilbert term albeit in
5D, a scalar field $\phi$ which has support in the bulk as well as on the
brane, and a matter Lagrangian density, ${\cal L}^{(i)}(\phi)$
on each of the branes, 
\beq
S=\frac{1}{2\kap} \int_M d^5x \sqrt{-g} \left\{ R - \half (\pd \phi)^2 -
                                                \kap V(\phi) \right\}
-\frac{1}{2\kap} \sum_i \int_{M_i} d^4x \sqrt{-h^{(i)}} 
\left\{ 2\jump{K^{(i)}} + \kap U^{(i)}(\phi) + \kap \lag^{(i)}(\phi)
\right\} \,,
\label{action}
\eeq 
where $M$ is the 5D manifold  with metric $g_{ab}$, and the 4D
hypersurfaces  $M_i$ are the branes each having a projected metric
$h^{(i)}_{ab}$.  The constant $\kap$ , which is effectively the 5D
gravitational constant, has mass dimensions of ${\cal O}(m^{-3})$
 ($m$ is a unit
of mass), $\phi$ is dimensionless, the bulk potential $V(\phi)$ has
dimensions ${\cal O}(m^5)$ and the brane potentials $U^{(i)}(\phi)$ have
dimensions ${\cal O}(m^4)$.  The $2\jump{K^{(i)}}$ terms added to action on the
brane  are the appropriate Gibbons-Hawking terms~\cite{GH}, required for a
well-formulated variational problem.

Varying this action with respect to the metric gives
\begin{eqnarray}
\nonumber \delta_g S&=&\frac{1}{2\kap} \int_M d^5 x \sqrt{-g} \left\{
G_{ab} - \half \pd_a \phi \pd_b \phi + \half g_{ab} \left( \half (\pd
\phi)^2 + \kap V(\phi) \right) \right\} \delta g^{ab} \\
&&{}+\frac{1}{2\kap} \sum_i \int_{M_i} d^4 x \sqrt{-h} \left\{
\jump{Kh_{ab} - K_{ab}} + \frac{\kap}{2} \left( U(\phi) h_{ab} + \lag
(\phi) h_{ab} - 2\varlag{g^{ab}} \right) \right\} \delta g^{ab} \,,
\label{metricvariation}
\end{eqnarray}
where we have now dropped the $i$ labels on the terms in the brane action
for convenience; suffice to say, all the subsequent expressions apply
on each of the branes.  The $\jump{Kh_{ab}-K_{ab}}$ term is a
combination of the boundary terms from the variation of the bulk
Einstein-Hilbert term and that  of the Gibbons-Hawking term
(see ref.~\cite{CR} for more details as to its derivation).  By demanding
these expressions be true for all variations, one
can read off the bulk Einstein equations and the
Darmois-Israel junction conditions\footnote{These are almost
universally referred to as
the Israel junction conditions~\cite{israel}, but this extrinsic curvature
method was used much earlier by the French mathematician 
George Darmois~\cite{darmois}. We would like to thank Brandon Carter
for pointing this out to us.  Even earlier work on junctions in
spacetimes was done by Lanczos~\cite{lanczos} and Sen~\cite{sen}.}
to be 
\begin{eqnarray}
\label{bulkem}
&&G_{ab}=\half \pd_a \phi \pd_b \phi - \half \left( \half (\pd \phi)^2 +
         \kap V(\phi) \right) g_{ab} \,,\\
&&\jump{Kh_{ab}-K_{ab}}=\kap \left(\tau_{ab}-\half
         U(\phi)h_{ab}\right) \,,
\label{dijunction}
\end{eqnarray}
where we have defined $\tau_{ab}$, which we will interpret as the
energy-momentum of `ordinary' brane matter, by
\beq
\tau_{ab}=\varlag{h^{ab}} - \half \lag h_{ab}
=(\sqrt{-h})^{-1}{\delta
\over\delta h^{ab}} \left(\sqrt{-h}\lag\right)\,.
\eeq

Varying (\ref{action}) with respect to $\phi$ gives
\beq
\delta_\phi S=\frac{1}{2\kap} \int_M d^5x \sqrt{-g} \left\{ \nabla^2 \phi -
                              \kap \frac{dV}{d \phi} \right\} \delta \phi
+\frac{1}{2\kap} \sum_i \int_{M_i} d^4x \sqrt{-h} \left\{ \jump{n.\pd\phi}
-\kap \frac{d U}{d \phi} - \kap \varlag{\phi} \right\} \delta \phi \,,
\label{phivariation}
\eeq
where $\nabla^2$ is the covariant 5D wave operator $g^{ab}\nabla_a\nabla_b$.
Again, the $i$ labels on brane terms have been dropped for convenience.
Reading off the $\phi$ equation of motion in the bulk and the
jump condition at each brane gives
\beq
\label{phimotion}
\nabla^2 \phi = \kap \frac{dV}{d \phi} \,,\qquad
\jump{n.\pd\phi} = \kap \left( \frac{dU}{d \phi} + \varlag{\phi}
\right) \,.
\eeq
By using (\ref{proj}) and an appropriate generalization, one can write
$\nabla^2 \phi$ in terms of the 4D covariant wave operator,
$\delbar^2$, acting on $\phi$ and normal derivatives of $\phi$ as follows
\beq
\nabla^2 \phi = \delbar^2 \phi + K n.\pd \phi + n. \pd (n.\pd \phi) =
\kap{dV\over d\phi}\,.
\eeq
This is expression is the result of the equivalent procedure used to
derive (\ref{gbar}) for the Einstein tensor. However, as we shall
discuss in the subsequent sections, it is impossible to write this
equation in terms of local quantities only defined on the branes due to our
lack of knowledge of $n. \pd (n.\pd \phi)$ on the brane. This is the
main obstacle to the derivation of a closed form solution when there
is a bulk scalar field in an
equivalent way to the RS type models.

\subsection{Effective 4D equations}

Having derived the appropriate bulk Einstein equations and equations
motion for $\phi$ along with their appropriate junction conditions, we
are now in a position to derive effective 4D equations by applying the
geometrical projection technology of the section~\ref{sec:geom}.
The first  step in this direction is to impose $Z_2$ symmetry which
relates quantities at each side of the brane and allows us
to evaluate $K_{ab}$ and $n.\pd\phi$ close to, but not on, the brane
using the junction conditions,
\beq
K_{ab}=-{1\over 2}\kap\bigg(\tau_{ab}-{1\over 3}h_{ab}\tau+{1\over
6}Uh_{ab}\bigg)\,,\qquad n.\pd\phi={1\over 2}\kap\bigg({dU\over
d\phi}+\varlag{\phi}\bigg)\,.
\label{z2jun}
\eeq
  
Substituting the expression for $K_{ab}$, and the bulk Einstein
equation  (\ref{bulkem}) into (\ref{gbar}) gives the effective 4D
Einstein equation,
\beq
\label{einstein}
\Gbar_{ab} = -\Lambar(\phi) h_{ab} + \frac{\kapp}{12}U(\phi) \tau_{ab}
+\frac{\kapp}{16} U'(\phi) \varlag{\phi} h_{ab} + \kapp \pi_{ab} +
\frac{\kapp}{32} \left( \varlag{\phi} \right)^2 h_{ab}
+ \frac{1}{3} \delbar_a \phi \delbar_b \phi 
- \frac{5}{24}(\delbar \phi)^2 h_{ab} - E_{ab} \,,
\eeq
where, for convenience, we have written
\begin{eqnarray}
\Lambar(\phi)&=&\frac{\kapp}{48} U(\phi)^2 + \frac{\kap}{4}V(\phi) -
                 \frac{\kapp}{32}U'(\phi)^2 \,, \\
\pi_{ab}&=&-\frac{1}{4}{\tau_a}^c\tau_{bc} + \frac{1}{12}\tau\tau_{ab}
           +\frac{1}{8}\tau^{cd}\tau_{cd}h_{ab} -
                 \frac{1}{24}\tau^2h_{ab} \,.
\end{eqnarray}
$\Lambda(\phi)$ is an effective cosmological constant type term,
$\pi_{ab}$ is the quadratic correction to the Einstein equations
first deduced, but interpreted somewhat differently, in
ref.~\cite{BDL1}, and $E_{ab}$ is the only non-local term, it being a
projection of the bulk Weyl tensor onto the brane. In the next section
we shall discuss how its effects can be understood simply in terms of
quantities which are local to the brane in the case where the metric
on the brane is spatially isotropic and homogeneous.

By taking the trace of (\ref{einstein}) we can obtain an expression for
the 4D Ricci scalar,
\beq
\label{ricci}
\Rbar=4\Lambar(\phi)-\frac{\kapp}{12}U(\phi) \tau
-\frac{\kapp}{4} U'(\phi) \varlag{\phi} + 
\frac{\kapp}{12} \left( \tau^2 - 3\tau^{cd}\tau_{cd} \right)
-\frac{\kapp}{8} \left( \varlag{\phi} \right)^2 +
\half (\delbar \phi)^2 \,, 
\eeq
which is independent of the contribution from the bulk Weyl tensor
$E_{ab}$ and will be useful when considering the expansion rate.

Similarly, substituting (\ref{z2jun})
into the Codacci equation (\ref{codacci})
on the brane gives us the following expression for the divergence of the brane energy-momentum tensor 
\beq
\label{energy}
\delbar^b \tau_{ab} = -\frac{1}{2} \delbar_a \phi \varlag{\phi} \,,
\eeq
which is an effective energy momentum conservation equation for the
matter on the brane. Interestingly, the brane energy-momentum is only
conserved either if $\lag$ is independent of $\phi$, or if $\phi$ is
constant on the brane.

One can also substitute (\ref{z2jun}) into the the equation of motion
for $\phi$ on the brane which gives,
\beq
\label{effectscalar}
\delbar^2 \phi + n. \pd (n.\pd \phi) = \kap V'(\phi) -
\frac{\kapp}{12} (\tau - 2U(\phi)) \left( U'(\phi) +\varlag{\phi}
\right)\,.
\eeq
We have written this in terms of a d'Alembertian wave operator on the
brane, local source terms, and the second derivative of $\phi$ in the direction
of normal to the brane. This term, which is analogous to  $E_{ab}$
in the case of the effective Einstein tensor, is the only term which
is dependent on the bulk. Unfortunately, in contrast to $E_{ab}$, 
 it is not possible to understand it simply in terms of quantities
local to the brane and hence in our subsequent discussion we will be
forced to make some assumptions as to the behaviour of $\phi$ on the brane.
In previous discussion we have pointed out that RS showed that the
theory of gravity on the brane was a tensor theory, and we see that in the
effective 4D Einstein equations are only slightly modified by the bulk
in the weak-field limit, but there is clearly no way of
thinking of the bulk scalar field in the same way; it being a truly 5D
quantity, and must be treated as such.

Now, motivated by the isotropy of the CMB and homogeneity of the
observed galaxy distribution, we choose the metric on the brane to be an
embedding of the 4D FRW universe, that is,  
\beq
h_{ab} dx^a dx^b = -dt^2 + a(t)^2 \gamma_{ij} dx^i dx^j\,,
\eeq
where  $\gamma_{ij}$ is the metric of the three-spaces of constant
curvature. In this case, standard results give
\begin{eqnarray}
\label{frw1}
\frac{1}{3} \Gbar_{00} &=&H^2+\frac{k}{a^2} \,, \\
\Rbar&=&6\left( \dot{H} + 2H^2 + \frac{k}{a^2} \right) \,,
\label{frw2}
\end{eqnarray}
where dots denote differentiation with respect to cosmic time, 
$a$ is the scale factor and $H=\dot{a}/a$ is the Hubble
parameter. Therefore, in order to derive equations for $H$, and hence an
analogy to the Friedmann equation, we have two alternatives: either
compute $E_{00}$ on the brane  and hence $\Gbar_{00}$ as advocated in
ref.~\cite{shiro}, or integrate
the equation for $H$ in terms of $\Rbar$ which implies the existence
of an integration constant. This exactly the equivalent approach to
those taken in refs.~\cite{BDL1,BDL2}. In the subsequent sections
on applications we shall perform these calculations explicitly for a
number of cases, to show that they are equivalent. For the rest of
this section we shall attempt to compare our results with particular
4D theories of gravity using the later of these two approaches and in
the next section we will discuss the precise dynamics of $E_{ab}$ on
the brane.

If we now assume that $\lag$ is independent of $\phi$ and that the
quadratic corrections are small, then
\beq
\Rbar=4\Lambar(\phi)-\frac{\kapp}{12}U(\phi) \tau+\half (\delbar
\phi)^2\,.
\eeq
The equivalent expression for Einstein gravity with a cosmological
constant is
\beq
\Gbar_{ab} = 8 \pi G_N \tau_{ab} - \Lambar h_{ab}\,,
\qquad\Rbar = 4\Lambar - 8 \pi G_N \tau  \,.
\eeq
Therefore, we see the directly the interpretation of $\Lambar$ as
a cosmological constant and if one defines the an effective
gravitational constant given by 
\beq
\label{Gnewton}
\frac{\kapp}{12}U(\phi) = 8 \pi G_N (\phi) \,,
\eeq
then one we will get the same expansion rate as Einstein gravity in
this limit.  As we have
already discussed, from the point of view of the brane we have no
control over the dynamics of the scalar field, but the fact that it is
related to the effective gravitational constant means that it must not
have changed much since the time of BBN in any realistic model of the
universe.  Note that in making this
definition we have justified the assumption of ignoring the quadratic
terms at late since they will be suppressed by a factor of $\tau/U$.

Clearly, by making these identifications we can make some interesting
general statements. First, we should note that the expression for the
Ricci scalar is independent of $E_{ab}$ and hence, if $\phi$ is
constant on the brane, it is completely
insensitive to anything which is not on the brane. This is not to say
that the gravitational dynamics are in general independent of the
bulk, just that in this specialized case  the only effect on the
effective Friedmann equation is via an integration
constant. Secondly, we see that $U(\phi)$ must be a strictly positive
number, at least after the epoch of BBN, in order to recover the
desired expansion rate. We will discuss this in detail in the context
of the Goldberger-Wise mechanism for stabilizing the extra dimension;
suffice to say if the effective tension on the brane is negative then
the corresponding expansion rate will also be negative.
This is  effectively rules out
the RS1 model for solving the hierarchy problem, even when there exists
a scalar field to stabilize the extra dimension. Finally, in such
models there is a simple and at first sights uncomplicated way to
ensure a zero effective cosmological constant on the brane, by setting
\beq U^{\prime}(\phi)^2-{2\over 3}U(\phi)^2={8\over\kap}V(\phi)\,,
\eeq an equation which can be solved analytically in the simple case
where $V(\phi)=0$ and $V(\phi)$ is a constant. This self-tuning
mechanism, first suggested in refs.~\cite{ADKS,KSS}, has been the subject
of a great deal of debate in the context of supergravity models where
$U$ and $V$ are derived from the same super-potential (see also~\cite{GNS,BCC}). Unfortunately,
it appears~\cite{Gubser} that no such supersymmetric model exists
without the a naked singularity somewhere in the bulk.

We have already noted that it was deduced in ref.~\cite{GT} that there
is effective BD gravity on the negative tension brane in the RS1 model
due to a localized mass distribution on the positive tension brane,
and this may at first sight appear to in conflict with our
results. However, here we are only considering the effect on expansion
in the spatially homogeneous and isotropic model, and hence their
assumptions do not apply. In fact, a localized mass distribution off
the brane cannot contribute the overall expansion rate of the brane
since its only gravitational effect must be via $E_{ab}$ which is
traceless and hence does not contribute to Ricci scalar.

Even though the analysis of ref.~\cite{GT} does not apply here, it is
sensible to compare our model to alternative theories of gravity such
as BD, rather than just Einstein gravity since it can help us to
constrain further the parameters of our model.  The field equations
for BD gravity in 4D without a cosmological constant are~\cite{mtw}
\beq
\label{bd}
\Gbar_{ab}=\frac{8\pi}{\varphi}\tau_{ab}+\frac{\omega}{\varphi^2}
\left[ \delbar_a \varphi \delbar_b \varphi - \frac{1}{2} h_{ab}
(\delbar \varphi)^2 \right] + \frac{1}{\varphi} \left[ \delbar_a
\delbar_b \varphi - h_{ab} \delbar^2 \varphi \right] \,, \eeq and the
BD field, $\varphi$, has equation of motion \beq
\label{bdfield}
\delbar^2 \varphi = \frac{8\pi}{3+2\omega} \tau \,.  \eeq Here, we
have kept the bars and $h_{ab}$ as the 4D metric to avoid ambiguity.
Taking the trace of (\ref{bd}) and substituting (\ref{bdfield}) gives
an expression for the Ricci scalar \beq
\label{bdricci}
\Rbar=-\frac{8 \pi}{\varphi}\left(\frac{2\omega}{3+2\omega} \right)
\tau + \frac{\omega}{\varphi^2}(\delbar \varphi)^2 \,, \eeq and hence
we make the following  identification \beq
\frac{\kapp}{12}U(\phi)=\frac{8\pi}{\varphi}\bigg(\frac{2\omega}{3+2\omega}\bigg)\,.
\eeq 
Thus, if $U(\phi)<0$ and  $\varphi>0$, the
Brans-Dicke parameter, $\omega$, must be in the range
$-3/2<\omega<0$.  Note that the same bounds on $w$ were found
in~\cite{GT} when considering localized mass distributions on the
negative tension brane.
Such values of $\omega$ can be ruled out by
experiment \cite{will}, reiterating that the RS1 model, and associated
models with a stabilized scalar field, are ruled out by experiment. 

\subsection{Conformal Transformations}
\label{sec:conf}

There has been considerable debate in the literature of scalar-tensor
gravity theories (such as Brans-Dicke theory) as to which frame
represents the physical one.
This issue is also pertinent in the case of dilatonic brane-worlds due
to the existence of a scalar field.
Scalar-tensor theories can be related by conformal transformations of
the metric, $g_{ab}\rightarrow\widetilde{g}_{ab}=\Omega(\phi)^2g_{ab}$
and it is well understood that changing to a different frame in a
cosmological solution can change the expansion rate (in particular, it
could change the results of the following section.
We have adopted the view of \cite{SS} that the frame in the
formulation of the action (\ref{action}) 
is the physically significant one in the sense that it is with respect
to this metric that free particles will follow geodesics.  As this
viewpoint is not universal, even in the 4D case
(see \cite{FGN} for a summary of different opinions)
it is interesting to study how our effective Einstein equations behave
under conformal transformations of the metric.
In particular, we can demonstrate that, unlike with the Jordan frame
of Brans-Dicke theory~\cite{SS,FGN}, the frame chosen by the induced
metric is the one in which the energy-momentum of the scalar field
does not contain second derivatives of $\phi$.
This means that any other frame would be philosophically problematic in that second
derivative terms necessarily give an energy-momentum which is not
positive-definite.
(Again, there are differences of opinion: not everyone considers
violations of positive-definiteness in the gravitational scalar sector
to be problematic~\cite{SS}!)

If the 4D induced metric transforms as $h_{ab} \rightarrow \widetilde{h}_{ab}=
\Omega(\phi)^2h_{ab}$ then the covariant derivatives which preserve
these metrics are related by, for example,
\beq
\widetilde{\nabla}_a v_b=\delbar_a v_b+\frac{\Omega'}{\Omega}
\left(h_{ab}h^{cd}\partial_d\phi-
\delta^c_{(a}\phi\partial_{b)}\phi\right) v_c \,,
\eeq
and the Einstein tensor transforms to (see~\cite{wald})
\beq
\widetilde{G}_{ab}=\Gbar_{ab}-2\frac{\Omega'}{\Omega}\delbar_a\delbar_b\phi
+\frac{2}{\Omega^2}\left(2{\Omega'}^2-\Omega''\right)
\delbar_a\phi\delbar_b\phi+\frac{2}{\Omega}\left(
\Omega'\delbar_c\delbar_d\phi+\Omega''\delbar_c\phi\delbar_d\phi\right)
h_{ab}h^{cd} \,.
\eeq
Note that $\widetilde{n}_a=\Omega n_a$ and $\widetilde{n}^a=\Omega^{-1} n^b$
and hence $E_{ab}$ must be conformally invariant as both ${C^a}_{bcd}$ and
$n^an_b$ are.  Likewise, ${\tau^a}_b$ and ${\pi^a}_b$ are
conformally invariant.  Substituting for $\Gbar_{ab}$ from
(\ref{einstein}) and expressing everything in terms of derivatives
relating to $\widetilde{h}_{ab}$, one can deduce the effective Einstein equation
in the new frame, which is
\begin{eqnarray}
\label{confG}
\widetilde{G}_{ab}&=&-\frac{\Lambar}{\Omega^2}\widetilde{h}_{ab}
+\frac{\kapp U}{12\Omega^2}\widetilde{\tau}_{ab}+\frac{\kapp U'}{16\Omega^2}
\frac{\kapp}{\Omega^2}\widetilde{\pi}_{ab}-\widetilde{E}_{ab}-
2\frac{\Omega'}{\Omega}\widetilde{\nabla}_a\widetilde{\nabla}_b\phi+
\left(\frac{1}{3}+2\frac{{\Omega'}^2}{\Omega^2}-2\frac{\Omega''}{\Omega}\right)
\widetilde{\nabla}_a\phi\widetilde{\nabla}_b\phi \nonumber \\
&&{}+\left\{\frac{\kapp U'}{16\Omega^2}\frac{\delta}{\delta\phi}\left(
\frac{\widetilde{\lag}}{\Omega^2}\right)+\frac{\kapp}{32\Omega^2}
\left(\frac{\delta}{\delta\phi}\left(\frac{\widetilde{\lag}}{\Omega^2}\right)\right)^2+
2\frac{\Omega'}{\Omega}\widetilde{\nabla}^2\phi+\left(2\frac{\Omega''}{\Omega}-
4\frac{{\Omega'}^2}{\Omega^2}-\frac{5}{24}\right)(\widetilde{\nabla}\phi)^2
\right\}\widetilde{h}_{ab} \,.
\label{confein}
\end{eqnarray}
Note that we can substitute for $\widetilde{\nabla}^2\phi$ using the
conformal transform of (\ref{effectscalar}) and that some of the terms
can be set to zero by suitable choice of $\Omega$.
However, the $\widetilde{\nabla}_a\widetilde{\nabla}_b\phi$
term will always be present unless $\Omega$ is constant, which, of course,
corresponds merely to a rescaling of units.  Thus the
original frame is the only one in which the terms corresponding to the
effective 4D energy-momentum of the scalar field can be positive
definite.  Also note that, in all frames, the coefficient of $\tau_{ab}$ has the same
sign as the vacuum energy, as in Brans-Dicke theory.  Frames other than the one we
have considered have an additional term proportional to $\tau
h_{ab}$; the coefficient of this term is constrained by gravitational
tests such as those carried out on solar system measurements.

In order to consider the expansion rate in the new frame can also
calculate the equation of motion for the transformed Ricci scalar,
which is
\beq
\widetilde{R}=4\frac{\Lambar}{\Omega^2}+\frac{\kapp U}{12\Omega^2}
\widetilde{\tau}+\frac{\kapp}{\Omega^2}\widetilde{\pi}+\frac{\kapp U'}{4\Omega^2}
\frac{\delta}{\delta\phi}\left(\frac{\widetilde{\lag}}{\Omega^2}\right)+
\frac{\kapp}{8\Omega^2}\left(\frac{\delta}{\delta\phi}\left(
\frac{\widetilde{\lag}}{\Omega^2}\right)\right)^2+6\frac{\Omega'}{\Omega}
\widetilde{\nabla}^2\phi+\left(6\frac{\Omega''}{\Omega}
+14\frac{{\Omega'}^2}{\Omega^2}-\frac{1}{2}\right)(\widetilde{\nabla}\phi)^2
\,. \eeq
As mentioned above, it is possible to substitute for
$\widetilde{\nabla}^2\phi$ using (\ref{effectscalar}), which gives
\beq
\widetilde{\nabla}^2\phi=2\frac{\Omega'}{\Omega}(\widetilde{\nabla}\phi)^2
-n.\pd(n.\pd\phi)+\kap V'(\phi)-\frac{\kapp}{12}(\widetilde{\tau}-2U(\phi))
\left(U'(\phi)+\frac{\delta}{\delta\phi}
\left(\frac{\widetilde{\lag}}{\Omega^2}\right)\right)\,.
\eeq
Thus we can remove the second brane derivatives in $\phi$ at the
expense of introducing the second derivative normal to the brane,
$n.\partial(n.\partial\phi)$.  This gives
\begin{eqnarray}
\label{confR}
\widetilde{R}&=&4\widetilde{\Lambda}+\left(\frac{\kapp U}{12\Omega^2}+
\frac{\kapp U'\Omega'}{2\Omega}\right)\widetilde{\tau}+\left(
\frac{\kapp U'}{4\Omega^2}+\frac{\kapp\Omega'U'}{\Omega}\right)
\left(\frac{\delta}{\delta\phi}\left(\frac{\widetilde{\lag}}{\Omega^2}\right)\right)+
\frac{\kapp}{\Omega^2}\widetilde{\pi}-\frac{\kapp\Omega'}{2\Omega}
\widetilde{\tau}\left(\frac{\delta}{\delta\phi}
\left(\frac{\widetilde{\lag}}{\Omega^2}\right)\right) \nonumber \\
&&{}+\frac{\kapp}{8\Omega^2}\left(\frac{\delta}{\delta\phi}
\left(\frac{\widetilde{\lag}}{\Omega^2}\right)\right)^2
+\left(6\frac{\Omega''}{\Omega}+2\frac{{\Omega'}^2}{\Omega^2}-
\frac{1}{2}\right)(\widetilde{\nabla}\phi)^2-6\frac{\Omega'}{\Omega}
n.\partial(n.\partial\phi) \,,
\end{eqnarray}
where we have defined
\beq
\widetilde{\Lambda}=\frac{\Lambar}{\Omega^2}+\frac{\Omega'}{4\Omega}
(6 \kap V'+\kapp UU')  \,.
\eeq
It can be seen that the coefficient of $\widetilde{\tau}$ in this frame is
different from the coefficient of $\tau$ in the original frame.
In particular, is is possible to choose $\Omega(\phi)$ so that the
expansion rate is positive even when the brane has negative tension.
However, the effect of the
$n.\partial(n.\partial\phi)$ term is impossible to quantify.  From the
expression for the Einstein tensor, it is possible to see that,
although one can change the sign of the dependence of the Hubble
parameter on the matter content by changing frame, one cannot change
the sign of the effective Newton constant.

Our original choice of frame for the 5D action was the 5D Einstein
frame and one might want to argue the case for other 5D frames.  The result
should be recoverable from our conformally transformed equation
(\ref{confein}).  This has been studied in~\cite{BV} starting from a
5D action in a general frame:
\beq
S=\frac{1}{2\kap}\int_Md^5x\sqrt{-g}\left\{F(\phi)R+\cdots\right\}
-\frac{1}{2\kap}\sum_i\int_{M_i}d^4x\sqrt{-h^{(i)}}
\left\{2F(\phi)\jump{K^{(i)}}+\cdots\right\} \,.
\eeq 
We refer the reader to~\cite{BV} for a discussion of the results and
merely quote that the effective Einstein equation derived from such an
action has the form
\beq
\Gbar_{ab}=8\pi G_N\tau_{ab}+8\pi G_A\tau h_{ab} + \cdots \,,
\eeq
where $G_N$ is in agreement with our conformally transformed equation
(\ref{confein}) above as expected.  In particular, it should be
noted that $G_N$ always has the same sign as the vacuum energy of the
brane.

In summary, although there are differences in opinion in the
literature of scalar-tensor gravity theories, the frame we have used
is in good agreement with the philosophy of both~\cite{SS}
and~\cite{FGN}, in the first case because we have interpreted the
metric of the original frame as that governing the geodesics and in the
second because ours is the only frame in which the scalar field has
positive definite energy-momentum.

\subsection{Understanding the effects of the bulk}

Our 4D effective equations for the gravitational sector
(\ref{einstein}) and the scalar field (\ref{effectscalar}) both
contain terms which specify how effects from the bulk can manifest
themselves on the brane. In the case of the scalar field it is
impossible to derive a simple expression in terms of just local
quantities forcing us to make sensible assumptions as to the behaviour
of $\phi$ on the brane. However, it is possible to do so for the in
the gravitational sector. 

One can deduce an equation for $E_{ab}$ on the brane
by realizing that the 4D Einstein
tensor must satisfy the 4D Bianchi identity $\delbar^a\Gbar_{ab}=0$.
From (\ref{einstein}) one can deduce that 
\begin{eqnarray}
\nonumber
\delbar^a E_{ab} &=& - \Lambar'(\phi)\delbar_b \phi +
 8\pi G'_N(\phi)  \tau_{ab}\delbar^a\phi
-4\pi G_N(\phi)  \varlag{\phi}\delbar_b \phi
+ \kapp \delbar^a \pi_{ab}
+ \frac{\kapp}{16} U''(\phi) \varlag{\phi} \delbar_b \phi \\
&&{}+\frac{\kapp}{16} \left( U'(\phi) + \varlag{\phi} \right)
\delbar_b \left( \varlag{\phi} \right)
+ \frac{1}{3} \delbar^2 \phi \delbar_b \phi
- \frac{1}{24} \delbar_b (\delbar \phi)^2 \,.
\label{bianchi}
\end{eqnarray}
The tensor $E_{ab}$ is traceless, tangential to the brane and
satisfies the above equation. Therefore, from the point of view of the
brane it includes 5 independent quantities, 3 of which are related to
spatial isotropy of the brane via $E_{0i}$ and the other 2 are tensor
degrees of freedom due to gravitational waves in the bulk.

If we assume that the brane and the scalar field on the brane
are spatially homogeneous and there are no
bulk gravitational waves then $E_{0i}=0$ since
$\Gbar_{0i}=0$, and  
$E_{ij}=f(t)\gamma_{ij}$ for some function $f(t)$. Therefore, $E_{ab}$
is entirely specified by the $b=0$ component of (\ref{bianchi}),
\begin{eqnarray}
\dot{E_{00}}+4\frac{\dot{a}}{a}E_{00} &=& \dot{\phi} \left\{
\Lambar'(\phi) + 8\pi G'_N(\phi)\tau_{00} + 4\pi G_N(\phi)
\varlag{\phi} -\frac{\kapp}{16}U''(\phi)\varlag{\phi} - \frac{1}{4}
\ddot{\phi} \right\} - \kapp \delbar^a \pi_{a0} \\
&&{}-\frac{\kapp}{16} \left\{ U'(\phi) + \varlag{\phi} \right\}
\frac{d}{dt} \left( \varlag{\phi} \right) \,.
\label{Eeqn1}
\end{eqnarray}
If $\lag$ is independent of $\phi$ and we consider a perfect fluid,
we can write
${\tau^a}_b = \diag (-\rho,p,p,p)$ and hence we find that 
$\delbar^a \pi_{a0} = 0$.
Thus
\beq
\label{Eeqn2}
\dot{E_{00}}+4\frac{\dot{a}}{a}E_{00} = \dot{\phi} \left\{
\Lambar'(\phi)+8\pi G'_N(\phi)\rho-\frac{1}{4} \ddot{\phi} \right\} \,,
\eeq
and so, if $\phi$ is constant,
\beq
\label{Eeqn3}
\frac{1}{a^4}\frac{d}{dt} \left( a^4 E_{00} \right) = 0 \,.
\eeq
which integrates to give $E_{00} = M a^{-4}$ where the integration
constant, $M$, has dimensions ${\cal O}(m^2)$. This is the exact same
integration constant as discussed in terms of integrating the equation
for the Ricci scalar in the previous section. Clearly, if $\phi$ is
slowly varying then modifications to this result can be derived in a
perturbative expansion.

We have shown, therefore, that the effect of the bulk on a spatially
isotropic and homogeneous brane is to contribute an effective
radiation type term to the Friedmann
equation~\cite{BDL2,Flan2,shiro,V,Muk}.  The physical
interpretation of the integration constant is that it quantifies the
mass outside the brane, which in the case of RS2 is zero and in RS1
is just that of the other brane. This is almost equivalent to Gauss'
law in electrostatics and is the analogue of Birkhoff's theorem for
spherically symmetric space-times (see~\cite{kraus} for a discussion
of Birkhoff's theorem applied to domain walls in adS space).
In fact this result is much deeper. In ref.~\cite{BCG} it was shown that the only global
spacetime compatible with spatial isotropy on the brane is
Schwarzschild-anti-de-Sitter and this integration constant is the
equivalent of the Schwarzschild mass term in usual treatment of
vacuum spacetimes in 4D.  Therefore, the precise
magnitude this contribution to the expansion rate not only depends on
the mass of other branes, but that of any black holes nucleated by
thermal effects in the bulk~\cite{GS,CKN}. Since it behaves like
radiation on the brane, the magnitude of such a contribution to the
expansion rate can be constrained using BBN.

A corollary of this, reiterating our earlier point, is that  when
$\dot\phi=0$, there is not possibility of changing the sign of the
effective gravitational constant due to effects from the bulk. Hence
when $\phi$ is stabilized a negative tension brane will have a
negative expansion rate. We should note that this does not preclude
such effects when the scalar field is varying, nor does it prevent
effects being transmitted between branes by perturbations from spatial
isotropy and homogeneity, for example, the localized mass distribution
of ref.~\cite{GT}. Such effects would, however, be at perturbative
order in the cosmological context.

\section{Applications}
\label{sec:apps}

\subsection{Randall-Sundrum type models}

The basic RS scenarios are a very special case of the general class of
models we have discussed where $\phi$ is a constant at all times. The
potentials will then take constant values, related to the CCs by \beq
V(\phi)= 2\Lambda \,, \qquad U_i(\phi)= 2\lambda_i \,, \eeq where
$\Lambda$ is the bulk CC and $\lambda_i$ are the brane CCs with
$i=1,2$ and $\lambda_1=-\lambda_2$ in RS1.  Substituting these
potentials into (\ref{einstein}) gives the result obtained in
ref.\cite{shiro}, that  \beq \Gbar_{ab}=\left( \frac{\kapp
\lambda^2}{12} + \frac{\kap \Lambda}{2} \right) h_{ab} + \frac{\kapp
\lambda}{6}\tau_{ab} + \kapp\pi_{ab} - E_{ab} \,, \eeq and the
standard energy conservation equation $\delbar^a \tau_{ab}=0$.  Note
that this result holds for both the RS scenarios: there will be an
equation for each brane in the first scenario, but that the  $i$
labels have been dropped for convenience.  From this we can deduce the
the only way in which information about the bulk can be communicated
to the brane is by the $E_{ab}$ term.  The first term, which is
the effective 4D cosmological constant, is usually set to zero by
tuning $\Lambda$, but we shall not assume this to be so here. Note,
once again, that recovering a positive expansion rate requires that
$\lambda>0$, as found in ref.\cite{shiro}, and on our brane, we
identify \beq G_N=\frac{\kapp \lambda}{48 \pi} \,.  \eeq

We can recover the Friedmann equations for  these solutions using our
formalism in two ways as described in the previous section. Using
(\ref{frw1}) and assuming that  the matter on the brane to be a
comoving fluid with energy-momentum tensor given  by
${\tau^{a}}_{b}=\diag(-\rho,p,p,p)$, one can deduce that  \beq H^2 =
\left( \frac{\kapp \lambda^2}{36} + \frac{\kap \Lambda}{6} \right) +
\frac{8\pi G_N}{3}\rho + \frac{\kapp}{36}\rho^2 - \frac{1}{3}E_{00}  -
\frac{k}{a^2} \,.  \eeq This is the now well known Friedmann equation
for a RS type model and includes a term analogous to that found in 4D,
a quadratic order correction and the often ignored unknown $E_{00}$.
When the brane metric is spatially isotropic and homogeneous,  we can
determine $E_{00}$ from  (\ref{Eeqn3}) which give  \beq
\label{friedmann}
H^2 = \left( \frac{\kapp \lambda^2}{36} + \frac{\kap \Lambda}{6}
    \right) + \frac{8\pi G_N}{3}\rho + \frac{\kapp}{36}\rho^2 +
    \frac{M}{a^4} - \frac{k}{a^2} \,.  \eeq

The other way to recover this solution is to use (\ref{frw2}), which
gives \beq
\label{diffH}
\dot{H}+2H^2=\left( \frac{\kapp \lambda^2}{18} + \frac{\kap
  \Lambda}{3} \right) +\frac{4\pi
  G_N}{3}(\rho-3p)-\frac{\kapp}{36}(\rho^2+3\rho p) - \frac{k}{a^2}
  \,.  \eeq in combination with the conservation of energy-momentum on
  the brane,  $\dot{\rho}+3H(\rho+p)=0$, which  can be used to remove
  $p$ in favour of $\rho$ and $\dot{\rho}$.  If the dependent variable
  is changed from $t$ to $a$, the differential equation for $H^2$
  derived in ref.~\cite{BDL2} is recovered and the solution
  (\ref{friedmann}) is once again found. In fact, this analysis is slightly
  more general than that of ref.~\cite{BDL2} as it only requires
  homogeneity and isotropy on the brane, not necessarily in the bulk.

One clear problem of these simple RS type models is the fine-tuning of
the CCs when one tunes $\Lambda$ to make the effective 4D
cosmological constant zero to get a Ricci-flat brane in the
absence of matter. It has been suggested that this unwanted aspect of the models
can be  removed by using a scalar field to stabilize the extra
dimension in an RS1 type model, and our approach has been set up to
deal with this kind of scenario.  Such a mechanism was proposed by
Goldberger and Wise~\cite{GW} with \beq V(\phi)=
\frac{\mu^2}{\kap}\phi^2 + 2\Lambda \,, \qquad U_i(\phi)=
\frac{\mu_i}{\kap}(\phi^2-\nu_i^2)^2 + 2\lambda_i \,, \eeq where the
$\mu$ and $\mu_i$ have dimensions ${\cal O}(m)$, while the $\nu_i$ are
dimensionless\footnote{We should note that our definitions are such
that the dimensions and the particular forms differ from those used in
ref.~\cite{GW}. However, the physics are exactly the same.}.  If the
field $\phi$ is varying, the equation for $E_{00}$ can no longer be
integrated exactly, nor can the analogue of (\ref{diffH}), both due to
the rather complicated dependence on $\phi$. The expansion rate will
in general be very different and hence the scalar field must stabilize
before BBN.  Not surprisingly, however, if the field is constant then
the whole situation just returns to that of a simple RS model and the
integration can once again be done exactly. If the field is slowly
varying then, at zeroth order, the solution will be that for $\phi$
constant with small perturbative  corrections. The model stabilizes
when the $\phi$ takes the values $\pm \nu_i$ on the branes.  Thus,
after stabilization, the coefficient of $\rho$ in the expression for
$H^2$ is the same as (\ref{friedmann}). Hence, if $\lambda_i$ is
negative then negative expansion is expected at late times as in the
simple RS model.

What we have deduced in a rather roundabout way --- adding the scalar
field and then it making constant after stabilization  --- is that the
scalar field can have no effect on the sign of the expansion rate once
it is stabilized; the reason being that, as (\ref{diffH}) comes from
the trace of (\ref{einstein}) and $E_{ab}$ is trace-free, there is no
contribution from the bulk Weyl tensor in (\ref{diffH}).  Thus the
only contribution to $H^2$ that can possibly arise from the bulk Weyl
tensor is that term arising from the integration constant when solving
(\ref{diffH}). This is the argument discussed in the previous section
on the general case applied to the specific case under consideration
here. In particular, we note that recovering the standard Hubble
expansion when $\rho$ is small requires that $\lambda$ be
positive\footnote{One could recover an expanding universe with
negative $\lambda$ during the radiation era by choosing $M$
sufficiently large to dominate the ordinary radiation,
but one would never get matter domination, since as soon as the
matter terms begin to dominate the negative expansion terms would kick
in and contraction would ensue. The process of expansion and
contraction would take place indefinitely in such a model.}.  The same
result was discovered in~\cite{BHLLM}, where an explicit 5D metric was
used.  From section~\ref{sec:conf} we see that it is
possible to recover a positive expansion rate for cosmological
solutions in the case of a negative tension brane, although this is
unsatisfactory as it does not give the usual Einstein equation and
leads to an indefinite contribution to the energy-momentum from the
scalar field.

This result is almost the total opposite of that obtained
in~\cite{csaki2,lesgourges,csaki3,cline2}.  In these papers,
the standard Friedmann equations are recovered from the RS1 scenario
by averaging over the bulk to obtain
an effective 4D theory in a similar way to the dimensional reduction
performed in Kaluza-Klein theories.  In simple terms, these authors make a specific
coordinate choice and integrate over one of the coordinates. 
To compare this to our covariant formalism of projection onto the
brane, we now attempt to formulate such an averaging for a general
manifold with co-dimension one branes without reference to any
particular coordinate system.

The covariant analogue of integrating over the bulk coordinate at each point on
the brane is to integrate along a congruence of curves.  Note that to
integrate tensors we must use a pull-back to bring them into the
tangent space at the point on the brane. 
This averaging procedure will depend on the congruence chosen and, in
the case where there is more than one brane, on the brane chosen.
For any brane, the natural choice for the congruence is clearly given by the geodesics
tangent to the normal on the brane.  This gives rise to the same
foliation discussed in section~\ref{sec:form}, which is well-defined near
the brane but may not defined throughout the bulk.  The most serious
objection to this in models with two branes is that the geodesics may
not be normal to the second brane and so a different result would be
obtained working from the other brane.  Note that the method of
projecting onto the brane only requires the foliation to be
well-defined in a neighbourhood of the brane.

This brane-world averaging should be contrasted with the usual
Kaluza-Klein dimensional reduction where there are no branes and 
all matter experiences all of the dimensions.
(Of course, our projection method is meaningless in theories such as
these where there is no preferred hypersurface.)
In such theories, everything is considered to be relatively
homogeneous across the extra dimensions, i.e., there are no preferred
points.  In contrast, with brane worlds the matter has support only on
the branes and, in addition, the background is usually highly warped.
The effect of integrating over the bulk with the exponential weight
factor is to give 4D quantities which are largely determined by the
matter on or close to the shadow brane.
It is questionable whether it is sensible to interpret the average
as physically meaningful, especially as our experiences relate to a
very atypical point in the spacetime.

As the Goldberger-Wise method for stabilising the radius involves
introducing a scalar field, it is possible to re-interpret the results
by a conformal rescaling, as discussed in section~\ref{sec:conf}.
Some authors, such as~\cite{csaki3}, use a conformal scaling chosen by
integrating over the bulk.  For the cosmological solutions they
consider, this gives a positive expansion rate, even for negative
tension branes: an effect which can be seen by considering the term
proportional to $\widetilde{\tau}$ in (\ref{confR}).  However, as we
have already observed from (\ref{confG}) this does not recover a
positive Newton constant in the full Einstein equations.

The difference in the results obtained by averaging and projecting
raises the philosophical issue of which should be interpreted as the
correct 4D effective theory.  This difference does not seems to have
been discussed in the literature, where some authors,
e.g.,~\cite{csaki2,lesgourges,csaki3,cline2},  have adopted the
averaging method whereas others, e.g.,~\cite{BDL1,BDL2,BHLLM}, have
obtained different results by hands-on versions of our method where a
specific form of the 5D metric is used.  In particular, \cite{BHLLM}
concludes, as we do, that the brane tension must be positive to get
the correct cosmological expansion even in presence of a stabilized
scalar field, whereas~\cite{csaki3} conclude the opposite by an averaging
analysis.  This discrepancy for cosmological models could be resolved
by reinterpreting the metric via a conformal scaling, but this is
problematic for reasons we have already discussed.  One technical advantage of the
projection procedure is that it can be defined on a larger class of
manifolds.  On a more philosophical level, the defining property of
brane-world models is that we are confined to the brane and so the
only matter we observe via non-gravitational interactions is that on
our brane and the curvature quantities we measure are those of the
metric restricted to the 4D submanifold, so the projection method
seems a more natural interpretation.

\subsection{Ho\v{r}ava-Witten model}

The Ho\v{r}ava-Witten model is a realistic scenario for a brane-world
based on the a compactification of 11D description of $E_8\times E_8$
heterotic string theory, often called heterotic M-theory~\cite{Ovrut}.
The general model contains many fields, most
of which can be set to zero self-consistently. However, one, a scalar
field, cannot since it corresponds to the deformation properties of the
compactified Calabi-Yau manifold which is used to reduce the model
from 11D to 5D. In this section we shall explore the implications of
similar models with a scalar field.  As a result of our analysis in
section~\ref{sec:conf}, we will work in the 4D conformal frame
selected by the induced metric.

Within the general framework in which we have been working, we shall
consider two branes and the potentials 
\beq
V(\phi)=\kappa_5^{-10/3} C e^{-2\beta\phi} \,, \qquad
U_i(\phi)=\kappa_5^{-8/3} C_i e^{-\beta\phi} \,,
\eeq
where the powers of $\kappa_5$ have been chosen to make the $C$ and $C_i$
dimensionless.
We expect these potentials to correspond to `slow-roll' behaviour of $\phi$.
Spherically symmetric cosmological solutions for dilaton domain walls with
these potentials have been found in ref.~\cite{CR} using Birkhoff's theorem.
The effective Einstein equation (\ref{einstein}) now has
\beq
\Lambar(\phi) = \frac{\kappa_5^{-4/3}}{96} e^{-2\beta\phi} 
 \left( 24C + (2-3\beta^2) C_i^2 \right) \,, \qquad
8\pi G_N (\phi) = \frac{\kappa_5^{4/3}}{12} C_i e^{-\beta\phi} \,,
\eeq
which leads to a variable cosmological constant term, similar to
quintessence, and also a variable effective gravitational constant.
From ref.~\cite{Ovrut} we see that, for the simplest HW theory, $\beta=1$ and
$24C=C_i^2$; thus $\Lambar=0$ in this model. We should note, however,
that the mechanism by which the a flat brane ($\Lambar=0$) is achieved
is subtly different from the RS type models since the potential in the
bulk is positive; the cancellation being due to spatial gradients of
the scalar field.
We also have $C_1=-C_2$ so, from the previous analysis, we must live
on the brane which has a positive value of $C_i$ to recover Einstein
gravity.   

We study the cosmology of this model using the same methods we
employed to the RS type models.  The equation for $H$ (\ref{frw2}) gives us the
following differential equation:
\beq
\label{HWdiffH}
\dot{H}+2H^2=\frac{\kappa_5^{4/3}}{72} C_i e^{-\beta\phi} (\rho-3p)
 - \frac{\kapp}{36}(\rho^2+3\rho p) - \frac{1}{12} \dot{\phi}^2 - 
   \frac{k}{a^2} \,,
\eeq

\noindent
where we have assumed, for simplicity, that $C$ is tuned in terms of
$C_i$ and $\beta$ so that the cosmological term vanishes
identically.
The equation for $E_{00}$ (\ref{Eeqn1}) becomes
\beq
\label{HWEde}
\dot{E_{00}}+4\frac{\dot{a}}{a}E_{00} = \frac{1}{12} \dot{\phi}
\left( 3 \ddot{\phi} - \kappa_5^{4/3} \beta C_i e^{-\beta\phi}\rho \right)
\,,
\eeq
and we have the standard energy-momentum conservation condition
$\dot\rho+3H(\rho+p)=0$. As we have already noted the effective
gravitational constant varies, but in the way which is different from
that usually assumed in 4D.  The standard way to allow $G_N$ to vary
in 4D GR is to make $\delbar^a(G_N\tau_{ab})=0$ as a
consequence of the Bianchi identities, but this is no longer true in
this case and $G_N$ can vary without a modification to 
energy-momentum conservation.  Hence this HW type model provides a
novel mechanism for the variation of $G_N$

At late times, we will require $\phi$ to be almost constant.  The solution
of zeroth order in $\phi$ is effectively the RS solution, but in a specialized
case we can attempt recover the next-to-leading order correction in
$\phi$. If we ignore the ${\dot{\phi}}^2$ term, the equation
(\ref{HWdiffH}) still cannot be integrated in general due to the
$\phi$ dependence in the first term.  However, if the matter satisfies
the equation of state $\rho=3p$, that is, if the universe is
radiation dominated, then the first term will be identically zero.
Integrating the energy conservation equation, gives the usual solution
for radiation, $\rho=\rho_0 a^{-4}$, where $\rho_0$ is the density of
radiation today, if we assume $a(t_0)=1$ and  $t_0$ is the time of the
present day.  Thus we can write (\ref{HWdiffH}) as \beq
a\frac{dH^2}{da}+4H^2 = -\frac{\kapp \rho_0^2}{9a^8} - \frac{2k}{a^2}
\,, \eeq which can be solved to give \beq
\label{HWH}
H^2 = \frac{\mathcal{C}}{a^4} + \frac{\kapp \rho_0^2}{36a^8} -
\frac{k}{a^2} \,.  \eeq where $\mathcal{C}$ is a constant of
integration. This integration constant corresponds to the combination
of the bulk effect and the effect of the radiation localized to the
brane.

We can also attempt to derive this solution by computing $E_{00}$.
With $\rho = \rho_0 a^{-4}$, (\ref{HWEde}) can be solved to first
order in derivatives of $\phi$;
\beq \frac{d}{dt} \left( a^4E_{00}
\right) = \frac{d}{dt} \left( \frac{\kappa_5^{4/3}}{12} C_i
e^{-\beta}{\phi} \rho_0 \right) \quad \Rightarrow \quad E_{00} =
\frac{1}{a^4} \left( \frac{\kappa_5^{4/3}}{12} \rho_0 C_i e^{-\beta\phi}
- 3 {\mathcal{C}} \right) \,, \eeq where we have chosen
$-3\mathcal{K}$ as the constant of integration for consistency with
(\ref{HWH}).  Equation (\ref{frw1}) gives us \beq
\label{HWH2}
H^2=\frac{\kappa_5^{4/3} \rho_0}{36a^4}C_i e^{-\beta\phi} + \frac{\kapp
\rho_0^2}{36a^8} - \frac{k}{a^2} - \frac{1}{3}E_{00} \,.  \eeq We note
that, when we substitute the solution for $E_{00}$, there is a
cancellation between the terms dependent on $\phi$ and the result is
consistent with (\ref{HWH}).  Thus we see that there are no
first-order corrections in derivatives of $\phi$ to the Friedmann
equation in the radiation dominated era,  There is, however, a $\phi$
dependent correction to $E_{00}$ which cancels with the term
proportional to $\rho$ in (\ref{HWH2}). We conclude, therefore, that
the RS type solution is valid if one just ignores term which are
second-order in derivatives.

This solution is valid during the radiation era and when $\phi$ is
rolling slowly enough for any second-order terms in derivatives of
$\phi$ to be neglected.  By the time the universe has reached the
matter domination, one might hope that the evolution of $\phi$ is now sufficiently slow
that $\phi$ will be almost constant, so that all
derivatives of $\phi$ may be neglected and hence the RS solution can
recovered in the matter era as well.  Thus one might imagine that 
the cosmological solutions
during matter-domination and during radiation-domination, except
possibly at very early times when $\phi$ is not changing slowly, are
those of the RS model. Given the very different nature of the HW and RS
models, this is somewhat remarkable.

\subsection{Cosmology with matter coupled to $\phi$}

Our setup allows us to study the case where the matter on the brane is
coupled to the scalar field $\phi$, i.e., when
$\delta \lag / \delta \phi \ne 0$.
We now study such a scenario in a cosmological context.  Since
(\ref{ricci}) has $\delta \lag / \delta \phi$ terms, we need an
explicit form of the Lagrangian.  Since we want a cosmological
solution, we want the energy-momentum tensor on the brane to be
$\tau_{ab} = F(\phi) \left( ph_{ab} + (\rho+p) U_a U_b \right)$:
$F(\phi)$ is an arbitrary function, $\rho$ is the energy
density, $p$ is the pressure and $U$ is the velocity of the flow.
This suggests that we try $F(\phi)$ times the Lagrangian for a perfect
fluid\footnote{See \cite{stewart} for details of how this is
constructed.}, so we consider the variation where the matter part of the action is
given by
\beq
\label{matteract}
S_{\mathrm{matter}}=-2\int \sqrt{-g}d^4x F(\phi) \left[ p(\eps,s) - 
\frac{n}{2\eps} (h^{ab} \Omega_a \Omega_b + \eps^2) \right] \,.
\eeq
Here, $s$ is the entropy, $\eps$ is the enthalpy, $n$ is a Lagrange
multiplier and $\Omega$ is a 4-vector field, which can be written as
$\Omega_a = \delbar_a \chi + \alpha \delbar_a \beta + \theta \delbar_a s$
for some scalar fields $\chi$, $\alpha$, $\beta$ and $\theta$:
respectively the three Clebsch potentials and the thermasy.
Varying (\ref{matteract}) with respect to $n$ gives the constraint
$h^{ab} \Omega_a \Omega_b + \eps^2 = 0$ so we define the velocity $U =
\Omega / \eps$.
Varying with respect to the other parameters $\eps$, $\chi$, $\alpha$,
$\beta$, $\theta$ and $s$ gives
\beq
\frac{\pd p}{\pd \eps} = n \,, \qquad
\delbar_a \left[ n F(\phi) U^a \right] = 0 \,, \qquad
U^a \delbar_a \beta = U^a \delbar_a \alpha = U^a \delbar_a s = 0 \,, \qquad
\frac{\pd p}{\pd s} + n U^a \delbar_a \theta = 0 \,.
\eeq
Varying with respect to the metric gives, as required,
\beq
\tau_{ab} = F(\phi) \left( ph_{ab} + n\eps U_a U_b \right)
          = F(\phi) \left( ph_{ab} + (\rho+p) U_a U_b \right) \,.
\eeq
Thus, on-shell where $U^aU_a=-1$, the Lagrangian and its variation
with respect to $\phi$
\beq
\lag(\phi) = -2 F(\phi) p(\eps,s) \,, \qquad
\varlag{\phi} = -2 F'(\phi) p(\eps,s) \,.
\eeq
From (\ref{einstein}), we see that the effect of coupling the matter
can be thought of as changing the relation between $\rho$ and $p$.
Couplings of scalar fields to photons and
baryons are tightly constrained; coupling to dark matter is less
constrained (see \cite{BM} and refs.\ therein for more details).
As before we have chosen to work in the frame determined by the metric
induced on the brane.  As discussed in section~\ref{sec:conf}, there
is an issue of which frame to choose.  An alternative to the above
approach would be to transform the action into the Jordan frame
before varying the action and then transform the equations of motion
back to the Einstein frame.  The variation of the action in arbitrary
frames is given in~\cite{BV}.

Now that we have the expression for the Lagrangian that we need, let
us turn our attention to the cosmology of such models.
If we assume that $V(\phi)$ has been tuned so that $\Lambar=0$ and that
$\phi$ is spatially homogeneous on the brane (i.e., $\phi$ on the
brane depends only on $t$)
then (\ref{ricci}) and (\ref{frw2}) give us
\beq
\label{couplediffH}
\dot{H} + 2H^2 = 
\frac{\kapp}{72} U(\phi) F(\phi) (\rho - 3p) + \frac{\kapp}{12}
U'(\phi) F'(\phi) p - \frac{\kapp}{36} F(\phi)^2 (\rho^2 +3\rho p) -
\frac{\kapp}{12} F'(\phi)^2 p^2 - \frac{1}{12} \dot{\phi}^2 -
\frac{k}{a^2} \,,
\eeq
and (\ref{energy}) gives us
\beq
\label{noncons}
\dot{\rho} + 3H(\rho+p) = -\frac{F'(\phi)}{F(\phi)} \, \dot{\phi} \,
(p+\rho) \,.
\eeq
For an equation of state $p=w\rho$, (\ref{noncons}) has solution
$\rho=\rho_0 \left( a^3 F \right) ^{-(1+w)}$.
For dust ($p=0$), substituting this solution into (\ref{couplediffH})
gives
\beq
\dot{H} + 2H^2 = 
\frac{\kapp}{72} U(\phi) \frac{\rho_0}{a^3} - \frac{\kapp \rho_0^2}{36a^6}
- \frac{1}{12} \dot{\phi}^2 - \frac{k}{a^2} \,,
\eeq
which is identical to the non-coupled case.  This is not surprising
as, the energy-momentum tensor is 
${\tau^a}_b=\diag(\rho_0 a^{-3},0,0,0)$
so we have the same matter content as for non-coupled dust.
It is also a promising start for this model as the standard Big-Bang
model has been very successful as a description of the
matter-dominated universe and so we want our model to look like the
standard cosmology during the matter era.

Now consider radiation matter in the specific case where $\phi$ is
constant. 
For convenience, we will assume that $\rho^2$ and $\dot{\phi}$ terms
can be neglected.
Then (\ref{couplediffH}) becomes
\beq
\frac{d}{dt}(a^4H^2) = A \frac{\dot{a}}{a} - 2ka\dot{a} \,,
\eeq
where $A$ is a collection of constants.
We get the solution
\beq
\label{coupleH}
H^2 = \frac{A \log a}{a^4} + \frac{\mathcal{C}}{a^4} -
\frac{k}{a^2} \,.
\eeq
For the non-coupled radiation matter, $A$ in
(\ref{coupleH}) is always zero but the coupling allows a non-zero,
positive value.
As $a$ grows, this new term will dominate the $a^{-4}$ term.
Also, as $a \rightarrow 0$, the first term will dominate and will be
negative.
Thus there will be a value of $a$, dependent on the values of $\rho_0$,
$\mathcal{C}$ and $k$, for which $H^2=0$.

The term proportional to $a^{-4}$ contains a contribution from the
bulk and one from radiation matter on the brane.  To calculate these
individual contributions, we can use (\ref{frw1}) which, ignoring
quadratic effects, gives
\beq
H^2=\frac{\kapp \rho_0}{48a^4} \left(4UF^{-1/3} + U'F'F^{-4/3} \right)
-\frac{1}{3} E_{00} + \frac{k}{a^2} \,.
\eeq
We can calculate $E_{00}$ from (\ref{Eeqn1}) which reduces to
\beq
\frac{d}{dt}(a^4 E_{00}) = \frac{\kapp w \rho_0 U' F'}{8 F^{4/3}}
\frac{d}{dt}(\log a)
\,,
\eeq
where quadratic matter terms have again been neglected.
This generates the $a^{-4} \log a$ term and the other part of the
$a^{-4}$ term in (\ref{coupleH}).

We have a Friedmann equation which is like that of the standard
cosmology but with the extra $ a^{-4} \log a$ term.  Motivated by this,
we now study the cosmology generated by a Friedmann equation
with this new term in addition to the the usual radiation, dust and
curvature terms.  Thus we consider
\beq
\label{toyH}
H^2 = \frac{A \log a}{a^4} + \frac{B}{a^4} + \frac{C}{a^3} -
\frac{k}{a^2} \,,
\eeq
where $A$, $B$ and $C$ are positive constants.  We will consider this
very much as a dynamical system, although we will make some physically
motivated assumptions about the relative magnitudes of the constants.

To determine whether this models suffers from a flatness problem, 
we define the curvature density in the usual way by
$\Omega_k = -k / (a^2H^2)$, whence we derive the relation,
\beq
\dot{\Omega}_k=-\frac{2}{H} \left( \dot{H}+H^2 \right) \Omega_k \,.
\eeq
This tells us that $\Omega_k=0$ is a stable fixed point if
$\dot{H}+H^2>0$.  From (\ref{toyH}), we get
\beq
\dot{H}+H^2=\left( \frac{A}{2} - B \right) \frac{1}{a^4} -
\frac{A\log a}{a^4} - \frac{3C}{2a^3} \,.
\eeq
So there will be value of $a$, dependent on $A$, $B$ and $C$, below
which $\Omega_k=0$ will be an attractor.
This stability condition can be written,
\beq
\label{acc:cond}
A - 2B - 2A \log a - 3Ca > 0 \,.
\eeq
The r.h.s.\ of (\ref{toyH}) can be negative for some values of $a$.
Clearly we must have $H^2 \ge 0$, so we also require that $a$ only
takes values for which the r.h.s.\ of (\ref{toyH}) is non-negative.
We will neglect the last term both because its relative effect is only
significant for large $a$ and because we hope to ensure that the
universe is driven towards flatness.  Thus we must also satisfy the
condition
\beq
\label{pos:cond}
A \log a + B + Ca > 0 \,.
\eeq
If we can satisfy both (\ref{acc:cond}) and (\ref{pos:cond})
we will get a period of
accelerated expansion which drive the universe towards flatness.
Adding twice the second inequality to the first gives us $A>Ca$.
Substituting $a=A/C$ into the second inequality gives
\beq
\label{nec}
\log \left( \frac{C}{A} \right) < 1 + \frac{B}{A} \,.
\eeq
If this is not satisfied, then (\ref{pos:cond}) cannot hold for any
value of $a<A/C$.
Thus, the coefficients $A$, $B$ and $C$ must satisfy (\ref{nec})
for us to get a period in the history of the universe where
$\Omega_k=0$ is an attractor.
Note that this is a necessary but not sufficient condition.
We now consider whether we can achieve a period of accelerated
expansion for some reasonable values of the parameters $A$, $B$ and $C$.
Define ${\mathbf{B}}=B/A$ and ${\mathbf{C}}=C/A$.
Then conditions
(\ref{acc:cond}) and (\ref{pos:cond}) become
\begin{eqnarray}
f_1(a) &\equiv& 1 - 2 {\mathbf{B}} - 2 \log a - 3 {\mathbf{C}} a > 0 \,,
\\
f_2(a) &\equiv& \log a + {\mathbf{B}} + {\mathbf{C}} a > 0 \,.
\end{eqnarray}
Now $f_1$ decreases monotonically from $f_1(0)=\infty$ whereas $f_2$
increases monotonically from $f_2(0)=-\infty$.
Thus $\exists \, \alpha>0$ such that $f_1(\alpha)=f_2(\alpha)$.
If $f_1(\alpha)=f_2(\alpha)>0$ then there is an interval $I \ni \alpha$
such that, $\forall a \in I$, $f_1(a)>0$ and $f_2(a)>0$.

The ratio $C/B = {\mathbf{C}} / {\mathbf{B}}$ is well known and, if
we set $a=1$ at present times, it is $10^4$ to good approximation \cite{KT}.
By plotting $f_1$ and $f_2$ against $a$ for different values of
${\mathbf{B}}$ we can determine that, for ${\mathbf{B}} > 11$, there
will be a period of accelerated expansion.
Choosing larger values of ${\mathbf{B}}$ makes this period occur for
smaller values of $a$.
In the standard cosmology, BBN occurs when $a \sim 10^{-11}$.
So we expect the period of accelerated expansion to finish before
this.
Choosing ${\mathbf{B}}=30$ gives a period of acceleration 
$10^{-13} \appleq a \appleq 1.5 \times 10^{-13}$, which would be
acceptable.
In practice, ${\mathbf{B}}$ is probably much larger than this as no
effect like the $a^{-4} \log a$ term has been observed.

As discussed above, there is a finite value of $a$ for which $H=0$.
Clearly, it is important that this is an unstable fixed point so that
the universe expands from this initial value.  Hence we now analyse
the stability of the fixed point. 
Changing (\ref{toyH}) to conformal time, we find
\beq
\left( \frac{da}{d\tau} \right)^2 = A \log a + B + Ca - ka^2 \,.
\eeq
Scaling $\tau \rightarrow \beta \tau$ and $a \rightarrow \gamma a$ gives
\beq
\frac{\gamma^2}{\beta^2} \left( \frac{da}{d\tau} \right)^2
= A \log a + ( B + A \log \gamma ) + C \gamma a - k \gamma^2 a^2 \,,
\eeq
so we can choose $\beta$ and $\gamma$ to give
\beq
\label{dynsys}
a'^2 = \left( \frac{da}{d\tau} \right)^2 = \log a + \hat{A} a -
\hat{B} a^2 \,, 
\eeq
where $\hat{A}$ and $\hat{B}$ are redefined constants with $\hat{A}$
being positive and $\hat{B}$ having the same sign as $k$.
Primes denote differentiation with respect to the rescaled conformal time.
If $k \le 0$ then $a'=0$ has exactly one root, whereas, if $k>0$,
the number of roots will be two, one or zero depending on the values
of $\hat{A}$ and $\hat{B}$.
The ratio of $\hat{A}$ to $\hat{B}$ is
\beq
\frac{\hat{A}}{\hat{B}} = \frac{C}{k} e^{B/A} \,.
\eeq
Since we expect $B \gg A$ for a reasonable cosmology, this ratio will
be large, so there will be two roots.
The case $k>0$ with fewer than two roots does not allow $a'^2>0$, so
does not produce any sort of cosmology.

To analyse the stability of this fixed point (the one with the smaller
value of $a$ in the case where there are two) we calculate $a''$
by differentiating (\ref{dynsys}).
We find
\beq
2a'' = a^{-1} + \hat{A} - 2 \hat{B} a \,.
\eeq
If $k \le 0$, $a'' >0 \; \forall a$.
If $k>0$ then $\exists \, \xi$ such that $a>\xi \Rightarrow a''<0$ and
$a<\xi \Rightarrow a''>0$.
However, $\hat{A} / \hat{B} \gg 1$ so $ \xi \gg 1$ and so we have
$a''>0$ at the fixed point with the smaller value of $a$.
Thus, in all cases, we have a cosmology where the universe expands
from a finite value of $a$.
In the case $k \le 0$, the universe will expand indefinitely whereas,
if $k>0$, the universe will expand to the other fixed point.  To
determine the stability of the other fixed point in the latter case,
note that
\beq
a''=\frac{1}{a} \left( \frac{1}{2} + \frac{\hat{A}}{2}a - \hat{B}a^2 \right)
   < \log a + \hat{A} - \hat{B} a^2 \quad {\mathrm{if}} \; a>e^{1/2} \,.
\eeq
Thus $a''<0$ at the second fixed point if $a>e^{1/2}$, which is likely as
$\hat{A} \gg \hat{B}$, so the universe will recollapse.

The upshot of of all this analysis is that we have models where the universe
expands from a finite value of $a$ with an early period of accelerated
expansion (i.e.\ where $\Omega_k=0$ is an attractor).
These cosmological scenarios are interesting as they avoid an initial
singularity and alleviate the flatness problem of the standard
cosmology.  In the above discussion, we have ignored the terms
quadratic in the matter density.  Thus the accelerated expansion must
occur at low enough energy scales for the quadratic effects to be
negligible but at high enough scales not to affect BBN.

\section{Conclusions}

Using the Gauss-Codacci formalism we have derived effective 4D
Einstein equations for a brane-world scenario with a bulk scalar
field.  By studying the trace of these equations we have been able to
circumvent our ignorance of the contribution from the bulk in the
gravitational sector, but not for the scalar field.  This allows us to
determine necessary properties for recovering the standard
cosmological expansion in the late time limit.  We have shown that the
two-brane RS scenario cannot be made consistent with the desired
expansion rate from standard 4D Einstein
gravity nor with observational constraints on Brans-Dicke gravity on
the negative tension brane.  This is true even if there is a
stabilization mechanism, that is, constant or very slowly changing
scalar field.  A necessary condition for the recovery of a positive
expansion rate is that the vacuum energy of the brane must be positive when
the scalar field has been stabilized.

The expansion rate of a cosmological solution can be changed in sign
by introducing a conformal scaling of the metric.
This is a highly contentious issue in the literature of scalar-tensor
gravity theories.
We have adopted the view that the induced metric must be that which
governs the geometry of the spacetime (and hence the Hubble parameter
of a cosmological solution).
In section~\ref{sec:conf}, we showed that, in any other frame, there
would be second derivatives of the scalar field into the Einstein
equation, resulting in an energy-momentum for the scalar field which
violates positive definiteness.
In addition, we see that, as with 4D Brans-Dicke gravity, the term in
the Einstein equations proportional to the energy-momentum tensor
still has a negative coefficient in the case of a negative tension
brane, the possible change in sign of the Hubble parameter being due
to the additional trace term. 
Dilatonic brane-world models are fundamentally different from 4D
scalar-tensor theories in that the choice of frame is really an issue
for the 5D theory, which then determines the frame of the 4D effective
theory.
In our calculation, we have started from a 5D Einstein frame and shown
that the induced metric determines a frame which is preferable according
to both of the criteria normally advanced in the study of 4D
scalar-tensor theories, namely that matter follows geodesics of the
metric and that the contribution to the Einstein equations from the
scalar field is positively definite.

Differences with results obtained by other authors who perform a
dimensional reduction of the 5D spacetime by integrating 
over the bulk lead us to examine the difference between interpreting
averaged or projected quantities as the relevant physical 4D entities.
This is potentially an important philosophical issue in the study of
higher dimensional spaces with warped geometries.
We have argued that the averaging is difficult to define in a unique,
coordinate-free way, whereas projection suffers from no such
problem. In particular, it must be correct in the case that we have
studied where the matter on the brane is localized by a
$\delta$-function, it may not necessarily be the case when the matter
is defined in terms of excitations of higher dimensional gravity as in
Kaluza-Klein models, for example. Clearly, this issue requires some
understanding of how the matter is localized onto the brane from a
fundamental point of view before it can be answered with any certainty.

For a model based on a HW type dilaton, we have been able to obtain a
next-to-leading order correction in $\dot{\phi}$ to the RS solution in
the case of a universe dominated by radiation.  The Friedmann equation
is the same as for the RS model due to a cancellation of $\phi$
dependent terms in the bulk contribution. Given that the HW type
models have a rather different mechanism for achieving a cancellation
of the effective cosmological constant, and that the asymptotic bulk
geometry is likely to be somewhat different, it might seem surprising
that the they reduce to the same effective Friedmann equation when the
scalar field is assumed to be slowly varying. This suggests that form
of the RS Friedmann equation is in fact universal in a brane-world
which has any chance of being compatible with the standard Big-Bang
model.

It is natural to consider the case where the brane matter is
non-minimally coupled to the scalar field.  We discover that this does
not change the cosmology when in a matter dominated era, but in the
case of relativistic matter there is an additional term in the
Friedmann equation.  This gives rise to a cosmology where the universe
expands from a finite value of the scale factor, with a period of
accelerated expansion in the early universe.  Although a toy model, it
has the advantages of avoiding the initial singularity and flatness
problem of the standard big bang model. Clearly, more should be
directed in toward understanding this novel scenario.

\section*{Acknowledgements}

We would like to thank B.\ Carter, A.\ Chamblin, D.\ Langlois, H.\ Reall,
T.\ Shiromizu, J.\ Stewart and T.\ Wiseman for helpful comments.
RAB is supported by a PPARC Advanced Research Fellowship.
We are aware of related but independent work by D.\ Wands and
K.\ Maeda~\cite{Wands}.

\def\jnl#1#2#3#4#5#6{\hang{#1, {\it #4\/} {\bf #5}, #6 (#2).} }
\def\jnltwo#1#2#3#4#5#6#7#8{\hang{#1, {\it #4\/} {\bf #5}, #6; {\it
ibid} {\bf #7} #8 (#2).} } \def\prep#1#2#3#4{\hang{#1, #4.} }
\def\proc#1#2#3#4#5#6{{#1 [#2], in {\it #4\/}, #5, eds.\ (#6).} }
\def\book#1#2#3#4{\hang{#1, {\it #3\/} (#4, #2).} }
\def\jnlerr#1#2#3#4#5#6#7#8{\hang{#1 [#2], {\it #4\/} {\bf #5}, #6.
{Erratum:} {\it #4\/} {\bf #7}, #8.} } \def\prl{Phys.\ Rev.\ Lett.}
\def\pr{Phys.\ Rev.}  \def\pl{Phys.\ Lett.}  \def\np{Nucl.\ Phys.}
\def\prp{Phys.\ Rep.}  \def\rmp{Rev.\ Mod.\ Phys.}  \def\cmp{Comm.\
Math.\ Phys.}  \def\mpl{Mod.\ Phys.\ Lett.}  \def\apj{Astrophys.\ J.}
\def\apjl{Ap.\ J.\ Lett.}  \def\aap{Astron.\ Ap.}  \def\cqg{Class.\
Quant.\ Grav.}  \def\grg{Gen.\ Rel.\ Grav.}  \def\mn{Mon.\ Not.\ Roy.\
Astro.\ Soc.}
\def\ptp{Prog.\ Theor.\ Phys.}  \def\jetp{Sov.\ Phys.\ JETP}
\def\jetpl{JETP Lett.}  \def\jmp{J.\ Math.\ Phys.}  \def\zpc{Z.\
Phys.\ C} \def\cupress{Cambridge University Press} \def\pup{Princeton
University Press} \def\wss{World Scientific, Singapore}
\def\oup{Oxford University Press}


\begin{thebibliography}{99}

\bibitem{RubS}
\jnl{V.A.\ Rubakov and M.\ Shaposhnikov}{1983}{}{\pl}{125B}{136} \\
\jnl{K Akama}{1982}{Pregeometry}{Lect.\ Notes Phys.}{176}{267--271}

\bibitem{dim}
\jnl{N.\ Arkani-Hamed, S.\ Dimopoulos and G.\ Dvali}{1998}{}{\pl}{429B}{263} \\
\jnl{I.\ Antoniadis, N.\ Arkani-Hamed, S.\ Dimopoulos and G.\ Dvali}{1998}{}{\pl}{436B}{257}

\bibitem{RS1}
\jnl{L.\ Randall and R.\ Sundrum}{1999}{A large mass hierarchy from a
small extra dimension}{\prl}{83}{3370--3373}

\bibitem{RS2}
\jnl{L.\ Randall and R.\ Sundrum}{1999}{An alternative to
compactification}{\prl}{83}{4690--4693}

\bibitem{early}
\jnl{M.\ Visser}{1985}{}{\pl}{159B}{22} \\
\jnl{E.J.\ Squires}{1986}{}{\pl}{167B}{286} \\
\prep{M.\ Gogberashvili}{1998}{}{hep-ph/9812296} \\
\jnl{M.\ Gogberashvili}{1998}{}{Europhys.\ Lett.}{49}{396--399} \\
\jnl{M.\ Gogberashvili}{1998}{}{Mod.\ Phys.\ Lett.}{A14}{2025--2032}

\bibitem{hw}
\jnl{P.\ Ho\v{r}ava and E.\ Witten}{1996}{Heterotic and type I string
dynamics from eleven dimensions}{\np}{B460}{506} \\
\jnl{P.\ Ho\v{r}ava and E.\ Witten}{1996}{Eleven dimensional
supergravity on a manifold with boundary}{\np}{B475}{94}

\bibitem{witten}
\jnl{E.\ Witten}{1996}{Strong coupling expansion of Calabi-Yau
compactification}{\np}{B471}{135}

\bibitem{Ovrut}
\jnl{A.\ Lukas, B.A.\ Ovrut, K.S.\ Stelle and D.\ Waldram}{1999}{The
Universe as a Domain Wall}{\pr}{D59}{086001} \\
\jnl{A.\ Lukas, B.A.\ Ovrut and D.\ Waldram}{1999}{Cosmological
Solutions of Ho\v{r}ava-Witten Theory}{\pr}{D60}{086001} \\
\jnl{A.\ Lukas, B.A.\ Ovrut, K.S.\ Stelle and
D.\ Waldram}{1999}{Heterotic M-theory in Five
Dimensions}{\np}{B552}{246} \\
\jnl{A.\ Lukas, B.A.\ Ovrut and D.\ Waldram}{2000}{Boundary
inflation}{\pr}{D61}{023506}

\bibitem{Harvey}
\jnl{H.\ Reall}{1999}{}{\pr}{D59}{103506}

\bibitem{GT}
\jnl{J.\ Garriga and T.\ Tanaka}{2000}{Gravity in the Randall-Sundrum
brane world}{\prl}{84}{2778}

\bibitem{BDL1}
\jnl{P.\ Bin\'{e}truy, C.\ Deffayet and D.\ Langlois}{2000}{Non-conventional 
cosmology from a brane-universe}{\np}{B565}{269--287}

\bibitem{CF}
\jnl{D.\ Chung and K.\ Freese}{2000}{}{\pr}{D61}{2000}

\bibitem{cline}
\jnl{J.M.\ Cline, C.\ Grojean and G.\ Servant}{1999}{}{\prl}{83}{4245}

\bibitem{csaki1}
\jnl{C.\ Cs\'{a}ki, M.\ Graesser, C.\ Kolda and
J.\ Terning}{1999}{Cosmology of one extra dimension with localized
gravity}{\pl}{B462}{34}

\bibitem{BDL2}
\jnl{P.\ Bin\'{e}truy, C.\ Deffayet, U.\ Ellwanger and
D.\ Langlois}{2000}{Brane cosmological evolution in a bulk with
cosmological constant}{\pl}{B477}{285}

\bibitem{Flan2}
\jnl{E.E.\ Flannagan. S.-H.\ Tye and I.\ Wasserman}{2000}{}{\pr}{D62}{044039}

\bibitem{shiro} 
\jnl{T.\ Shiromizu, K.\ Maeda and M.\ Sasaki}{2000}{The Einstein
equations on the 3-brane world}{\pr}{D62}{024012}

\bibitem{csaki2} 
\jnl{C.\ Cs\'{a}ki, M.\ Graesser, L.\ Randall and
J.\ Terning}{2000}{Cosmology of brane models with radion
stabilisation}{\pr}{D62}{045015}

\bibitem{lesgourges}
\jnl{J.\ Lesgourges, S.\ Pastor, M.\ Pelso and L.\ Sorbo}{}{2000}{\pl}{B489}{411}

\bibitem{csaki3}
\jnl{C.\ Cs\'{a}ki, M.L.\ Graesser and G.D.\ Kribs}{2001}{}{\pr}{D63}{065002}

\bibitem{cline2}
\jnl{J.M.\ Cline and H.\ Firouzjahl}{2000}{}{\pl}{B495}{271--276}


\bibitem{GW}
\jnl{W.D.\ Goldberger and M.B.\ Wise}{1999}{Modulus stabilization with
bulk fields}{\prl}{83}{4922} \\
\jnl{W.D.\ Goldberger and M.B.\ Wise}{1999}{}{\pr}{D60}{107505}

\bibitem{V}
\jnl{D.\ Vollick}{2000}{}{\cqg}{18}{1--10}

\bibitem{Muk}
\jnl{S.\ Mukohyama}{2000}{}{\pr}{D62}{044039}

\bibitem{CR}
\jnl{H.A.\ Chamblin and H.S.\ Reall}{1999}{Dynamic dilaton domain walls}
{\np}{B562}{133--157}

\bibitem{Ull}
\prep{U.\ Ellwanger}{1999}{}{hep-th/0001126}

\bibitem{KOP}
\jnl{P.\ Kanti, K.A.\ Olive and M.\ Pospelov}{2000}{}{\prl}{85}{1374--1377}

\bibitem{wald}
\book{R.M.\ Wald}{1984}{General relativity}{University of Chicago Press}

\bibitem{mtw}
\book{C.W.\ Misner, K.S.\ Thorne and J.A.\ Wheeler}{1973}{Gravitation}{W.H.\ Freeman and Company}

\bibitem{CGR}
\jnl{C.\ Charmousis, R.\ Gregory and V.A.\ Rubakov}{2000}{}{\pr}{D62}{067505}

\bibitem{MSM}
\jnlerr{S.\ Mukohyama, T.\ Shiromizu and K.\ Maeda}{2000}{}{\pr}{D62}{024028}{D63}{024901}

\bibitem{GH}
\jnl{G.W.\ Gibbons and S.W.\ Hawking}{1977}{}{\pr}{D15}{2752}

\bibitem{israel}
\jnlerr{W.\ Israel}{1966}{Singular hypersurfaces and thin shells in
general relativity}{Nuovo Cimento}{44B}{1}{48B}{463}

\bibitem{darmois}
\jnl{G.\ Darmois}{1927}{Les \'{e}quations de la gravitation
einsteinienne}{M\'{e}morial des sciences math\'{e}matiques XXV}{}{}

\bibitem{lanczos}
\jnl{K.\ Lanczos}{1922}{}{unpublished}{}{} ---
\jnl{published}{1924}{}{Ann.\ Phys.\ (Leipzig)}{74}{518--540}

\bibitem{sen}
\jnl{N.\ Sen}{1924}{}{Ann.\ Phys.\ (Leipzig)}{73}{365--396}

\bibitem{ADKS}
\jnl{N.\ Arkani-Hamed, S.\ Dimopoulos, N.\ Kaloper and
R.\ Sundrum}{2000}{}{\pl}{B480}{193--199}

\bibitem{KSS}
\jnl{S.\ Kachru, M.\ Schulz and E.\ Silverstein}{2000}{}{\pr}{D62}{045021}

\bibitem{GNS}
\jnl{B.\ Grindstein, D.R.\ Nolte and W.\ Skiba}{2000}{}{\pr}{D62}{086006}

\bibitem{BCC}
\jnl{P.\ Bin\'{e}truy, J.M.\ Cline and C.\ Grojean}{2000}{}{\pl}{B489}{403--410}

\bibitem{Gubser}
\prep{S.\ Gubser}{2000}{}{hep-th/0002160}

\bibitem{BD}
\jnl{C.\ Brans and R.H.\ Dicke}{1961}{}{\pr}{24}{925}

\bibitem{will}
\prep{C.M.\ Will}{1998}{The confrontation between general relativity and experiment}{gr-qc/9811036}


\bibitem{kraus}
\jnl{P.\ Kraus}{1999}{}{JHEP}{9912}{001}

\bibitem{BCG}
\jnl{P.\ Bowcock, C.\ Charmousis and R.\ Gregory}{2000}{}{\cqg}{17}{4745--4764}

\bibitem{GS}
\jnl{J.\ Garriga and M.\ Sasaki}{2000}{}{\pr}{D62}{043523}

\bibitem{CKN}
\prep{A.\ Chamblin, A.\ Karch and A.\ Neyeri}{2000}{}{hep-th/0007060}

\bibitem{SS}
\jnl{D.I.\ Santiago and A.S.\ Silbergleit}{2000}{}{Gen.\ Rel.\ Grav.}{32}{565--581}

\bibitem{FGN}
\jnl{V.\ Faraoni, E.\ Gunzig and P.\ Nardone}{1999}{}{Fund.\ Cosmic Phys.}{20}{121}

\bibitem{BV}
\jnl{C.\ Barcel\'{o} and M.\ Visser}{2000}{}{JHEP}{0010}{019}

\bibitem{BHLLM}
\jnl{V.\ Barger, T.\ Han, T.\ Li, J.D.\ Lykken and D.\ Marfatia}{2000}{}{\pl}{B488}{97--107}


\bibitem{stewart}
\jnl{D.S.\ Salopek and J.M.\ Stewart}{1992}{Hamilton-Jacobi theory for
general relativity with matter fields}{\cqg}{9}{1943-1967}

\bibitem{BM}
\prep{R.\ Bean and J.\ Magueijo}{2000}{}{astro-ph/0007199}

\bibitem{KT}
\book{E.W.\ Kolb and M.S.\ Turner}{1990}{The Early Universe}{Addison-Wesley}

\bibitem{Wands}
\prep{K.\ Maeda and D.\ Wands}{2000}{}{hep-th/0008188}

\end{thebibliography}
\end{document}